\documentclass[12pt]{article}
\usepackage{amssymb}
\usepackage{amsmath}
\usepackage[dvips]{graphicx}
\usepackage[usenames]{color}
\usepackage{colortbl}
\newcommand{\ve}[1]{\boldsymbol{#1}}
\newcommand{\epsn}{\epsilon_N}

\newcommand{\en}{e_N}
\newcommand{\einf}{e_\infty}
\title{Numerical study of ground state energy fluctuations in spin glasses }
\author{Claudio Giberti$^a$ and Cecilia Vernia$^b$ \\
        \small $^a$ Dipartimento di Scienze e Metodi dell'Ingegneria, Universit\`a di Modena\\
        \small e Reggio Emilia, via G. Amendola 2 - Pad. Morselli, 42100 Reggio Emilia, Italy\\
        \small {\it claudio.giberti@unimore.it}\\
        \small $^b$ Dipartimento di Matematica Pura ed Applicata, Universit\`{a} di Modena\\
        \small e Reggio Emilia, via Campi 213/B, 41100 Modena, Italy\\
        \small {\it cecilia.vernia@unimore.it}}
\begin{document}
\maketitle
\begin{abstract}
\noindent Using a stochastic algorithm introduced in a previous
paper, we study the finite size volume corrections and the
fluctuations of the ground state energy in the
Sherrington-Kirkpatrick  and the Edwards-Anderson models at zero
temperature. The algorithm is based on a suitable annealing
procedure coupled with a balanced greedy-reluctant strategy that
drives the systems towards the deepest minimum of the energy
function.
\end{abstract}

\section{Introduction}
Finding solutions of computationally hard problems is an outstanding issue in applied science.
A classical example of  hard problem  is the search of the optimal configurations of a
functional with many local minima and a complex landscape: in fact, in this case the solution
is usually achieved with a computational effort that increases exponentially with the dimension
of the problem. Typically, such functionals arise in the modelling
of competing interactions among the components of large system.
In some interesting cases in physics and in other field (for example in biology, in economy,
in computer sciences,...), the interactions depend on some frozen-in structural disorder \cite{MPV},
and any realization of the disorder defines a sample of the system. Obviously,
the thermodynamic properties of such systems, like
spin glasses, fluctuate from sample to sample. Indeed,
the statistical variables of the system (spins, in the case of spin glasses)
interact via some random potential
which models the quenched disorder.
In this paper we consider the fluctuations, with respect to disorder realizations, of the low-lying states energy
of spin glasses (thus the functional to be optimized
here is the energy of the configurations of spins). Disorder-induced
fluctuations are particularly relevant for the physics of systems
at low temperatures. In fact, the properties of such systems are largely
dominated by the states with minimal energy into which the Gibbs-Boltzmann
distribution collapses as the temperature is decreased.

In order to be more specific, let us consider a disordered model with finite volume $N$ and described by a
Hamiltonian (or energy function) $H_N(J,\ve{S})$
which depends on a set of statistical variables $\ve{S}$  (the $N$ degrees of freedom of the system)
and on a set of variables $J$, representing the quenched disorder of the system,
which are randomly sampled according to
a given probability distribution.
For a fixed disorder realization $J$ and at zero temperature, the relevant states of the system
are only the ground states, that is the configurations $\ve{S}$ whose energy (for degree of
freedom) is given by $e_N=\frac 1N \min_{\ve{S}}H_N(J,\ve{S})$.
Being the minimum of a set of random variables, the energy density $\en$ is also a random variable
whose  probability density function (PDF)  $p_N(\en)$ follows some (a priori unknown) extreme values statistics:
we are interested in the numerical study of this distribution in its large volume limit.
In particular, we will mainly focus on the mean value $\epsn=\langle\en\rangle$ and on the standard deviation
$\sigma_N=(\langle\en^2\rangle-\langle\en\rangle^2)^{1/2}$ (here $\langle\ldots\rangle$ represents the average
with respect to disorder), but we will also try to give some insight into the shape of $p_N(\en)$.
In order to study their scaling behavior, we assume that $\epsn$ and $\sigma_N$  have a definite
limit as $N$ goes to infinity: $\epsn \to \einf$ and $\sigma_N\to 0$; in other words  we suppose that extensivity
 and self-averaging of the energy density $\en$ hold (these properties are, in fact, satisfied in  the models that we
are going to study). More precisely, assuming a power-law scaling behavior for the relevant quantities:

\begin{equation}
\epsn=\einf+bN^{-\omega}+\cdots, \qquad \qquad \sigma_N=aN^{-\rho}+\cdots,\label{e:ScalEn};
\end{equation}

we aim at estimating the exponents $\omega$ and $\rho$.\par\noindent
Since extensivity and self-averaging  imply a
trivial distribution  in the large volume limit ($p_N(x)\to \delta(x-\einf)$), we inquire into the
scaling behavior of $p_N(x)$ (if any) by studying the centered and scaled variable
$x_N=(\en-\epsn)/\sigma_N$ and seeking for its distribution $p_{\infty}(x)$ in the infinite volume limit.
\par
The previous issues have been numerically addressed by several authors in recent years \cite{BKM,P}
and a variety of spin glasses models have been considered \cite{B,B2,LPHJ}. All these studies call for large
computational resources. Indeed, the computational effort required to solve the intrinsically complex
problem of finding the ground state of $H_N$ \cite{MPV}, has to be multiplied by the large number of
realizations requested in the computation of the disorder averages over systems with larger and larger volumes.
In the cited papers \cite{BKM}-\cite{LPHJ} many different techniques are employed
in the approximation of the ground state.
Indeed, the relevance
of solving intrinsically complex problems,
has prompted many authors to devise efficient algorithms implementing many different approaches to this issue.
In this study we rely on a stochastic algorithm,  inspired by the classical
simulated annealing technique, which was introduced in \cite{CGGV2}. In that paper
the algorithm was challenged against the search of the ground state $e_N$ in
the Sherrington-Kirkpatrick model (SK), which
has become the standard of NP-complete problems.
Our purpose here is to push forward the validation of the algorithm by studying in some details the probability
distribution $p_N(e_N)$.
Moreover, forcing the scope of the algorithm that was optimized in \cite{CGGV2} for SK, we extend the study to a
preliminary analysis of the ground state of Edwards-Anderson model (EA). All the results that we are going to
describe are completely consistent with ones already present in the literature. \par
The paper is organized as follows. In section 2 we  introduce the
spin glasses studied in this work, namely the
Sherrington-Kirkpatrick  and the Edwards-Anderson models, and review
briefly some theoretical results that are relevant to the present
issues. We sketch the algorithm in section 3, referring  to the
original paper for details. The results are given in sections 4.

\section{The models}
The Ising spin glass models we study are the Sherrington-Kirkpatrick model (SK) \cite{SK},
whose Hamiltonian is defined by
\begin{equation}
H_N(J,\ve{S})= - \sum_{i,k=1 \atop i < k}^N J_{ik} S_i S_k,
\end{equation}
and the Edwards-Anderson model (EA) \cite{EA}, whose the Hamiltonian
is
\begin{equation}
H_N(J,\ve{S})=-\sum_{<i k>} J_{ik}S_iS_k,\label{e:EA}
\end{equation}
where $S_i=\pm 1$ for $i=1,\ldots ,N$ are Ising spin variables
which interact through couplings $J_{ik}$.
For both models, $J_{ik}$ are independent identically distributed
gaussian random variables,  with zero mean and variance $\frac 1N$ for
SK and $1$ for EA. With this choice of the variance the extensivity
and self-averaging of the ground state energy density of the two models
are guaranteed.

For three dimensional EA, which is a short range spin glass model,
the sum in (\ref{e:EA}) is over nearest neighbor spins on a given square
or cubic lattice of linear size $L$ (thus the number of nodes is $N=L^d$,
where $d$ is the dimension of the lattice), while for SK, which
is the mean field approximation of EA, the sum
is over all the spins. (For the physical origin of these models and a general
overview on spin glass, see \cite{BiYo})

The SK model in the low temperature phase was solved through the replica symmetry
breaking ansatz (RSB) by G. Parisi \cite{MPV}.
This solution, which is universally believed to be true for SK, is still a debated
issue in the mathematical physics community, because rigorous proofs of some fundamental
properties of the RSB scenario (for instance ultrametricity \cite{MPV}) are lacking.
However, recently some progress has been done in this direction with the mathematical proof of the Parisi
formula for the free energy density \cite{G,T}.
In the  framework of the RSB theory, the knowledge of the ground state energy in the thermodynamic limit
($\einf^P=-.7633..$ \cite{MPV}) and the probability distribution of large deviations of the free energy
at all temperatures \cite{PR} are available, but there are no exact analytic results
for the limiting behavior of the PDF.
A computation performed at the critical temperature $T_c$ and at the de Almeida-Thouless line in \cite{PRS},
gives a scaling exponent of the internal energy density equal to $\frac 2 3$.
In a more recent paper \cite{ABMM} a heuristic argument is presented which leads the same
value just below  $T_c$. It is natural to extrapolate this result to zero temperature obtaining for the
ground state energy density $\omega=\frac 2 3$. Indeed in \cite{ABMM} the authors argue, on the basis of the numerical
evidence, that this should be the scaling exponent in the whole spin glass phase. Similarly,
there are no analytical calculations for the sample-to-sample fluctuation exponent of the internal energy at
zero temperature. However, different analytical estimates for the fluctuation exponent of the free energy
density are available: $\frac 5 6$ \cite{PR,ABMM,CPSV} or $\frac 3 4$ \cite{AMY}.
As in the case of $\omega$, we can expect the exponent $\rho$ to be the same as that for the free energy density
in the whole region below $T_c$ \cite{BKM}, and in particular at zero temperature.
Unfortunately, given the closeness of the predicted values,
any numerical test at zero  as well as at finite temperature can hardly provide a
sharp distinction between two possibilities.

For the Edwards Anderson model things are even messier because there is not a general consensus on the nature
of its low temperature phase, and several scenarios (besides RSB) are conflicting \cite{NS}.
Regarding our concern, a relevant result was proven by Wehr and Aizenman \cite{WA}: they proved that
for EA, as for any finite-range spin glass model in finite dimension $d$, the scaling exponent for
the standard deviation $\sigma_N$ is $\rho=\frac 12$, namely that the ground-state energy variance
grows linearly with the volume $N=L^d$. In fact, in \cite{WA} it is shown that the variance
of {\it any} extensive quantity $\Psi_N$ depending on random parameters is of the order of the volume $N$.
Even if this result seems to suggest a gaussian limiting behavior for $(\Psi_N-\langle\Psi_N\rangle)/\sqrt{N}$,
the authors
of \cite{WA} state that this, in general, should be false, though only very mild violations to the normal distribution
are to be expected. However, for EA, heuristic arguments as well as numerical evidence point
to a gaussian limit for $p_N(e_N)$ \cite{BKM}. The same distribution and  the same estimate $\rho = \frac 12$
have been obtained by Aspelmeier and Moore with a replica theory calculation in \cite{AspMo}.

\section{The algorithm}

The numerical approach to the questions presented in Sec.1 requires the
computation of the minima of $H_N$ for a large sample of disorder realizations,
that is the search for the spin configuration which minimizes the energy (ground state configuration).
The minimization of a certain function depending on many discrete
variables (Hamiltonian) is a combinatorial optimization problem. Since often many combinatorial optimization
problems are NP complete, they are tackled by constructing approximation algorithms,
that run in acceptable amounts of computational time and have the property
that final configurations are ``close'' to globally minimal ones
(the metastable states approximate the ground state).
The models we consider are often presented as the standard example of NP-problems.
Indeed, the random sign (and strength) of the interaction
generates frustration in the systems, i.e. the fact
that in low energy configurations some of the couples
will have unsatisfied interaction. Therefore, the global minimization can not
be achieved simply by minimizing each local spin-to-spin interaction.
As a consequence, the ground state of the
systems is far from the standard ground state of
ferromagnetic models, where all spins point in the
same direction.

Several numerical studies have tried different
algorithms in the search of ground-state
energies, for example gradient descendent \cite{BP,CMPP,CGGUV1},
simulated annealing \cite{KGV,GSL},
genetic algorithms \cite{BKM,P}, branch-and-bound algorithm
\cite{KO} and extremal optimization \cite{B,BP01,BS}.
In this paper we use a stochastic algorithm which is the
optimal one in a class introduced in \cite{CGGV2}.
The algorithm is required to reach the lowest possible minima as quickly as
possible avoiding to get stuck in a local
minimum which is still far from the deepest ones.
In fact, in these complex systems disorder and frustration produce an energy
function with a corrugated landscape,
with a high multiplicity
of  valleys (local minima) separated by high barriers (local maxima).

Our algorithm generates a one spin-flip dynamics in the
configuration space; indeed, starting from a randomly chosen
initial condition, the algorithm explores the space through a
sequence of configurations obtained by inverting a single spin
sign to pass from a spin configuration to successive one.
The stochastic transition
from a trajectory point to another is ruled by a probability with an exponential density.

Essentially, the idea is that at each step one generates a priori a random energy jump (variation)
$\Delta H$ with probability $f_t(\Delta H)$ and then one moves in the direction
that produces the nearest energy variation to the chosen jump. The algorithm
stops in a local minimum that represents the best (sufficiently
deep) encountered minimum. This algorithm is justified by statistical properties of
the metastable states: these states are organized with a certain structure
 so that the dynamic evolution can be considered as the
overlapping of a fast motion in the attraction basin of a local minimum
and of a slow one with possibility of jumps between minima.
The possibility of exceed the energy barriers, needed to escape from poor local minima, is obtained by introducing
a sort of ``external temperature'' in the system, which
enables random positive energy fluctuations. The energy transitions
are generated in accordance with the following PDF

$$
f_t(x)=\left\{ \begin{array}{ll}
   e^{\lambda_1(t) x} & \textrm{se } x\leq 0\\
   e^{-\lambda_2(t) x}  & \textrm{se }  x > 0
\end{array}\right. , \qquad
\frac{1}{\lambda_1(t)} + \frac{1}{\lambda_2(t)}=1,
$$
in which the control parameter is mainly represented by
the cooling rate $\lambda_2(t)$ of the system and $t$ is the time of the dynamics, i.e.
the number of spin-flips since the beginning of the algorithm execution.
The choice of the exponential distribution is standard in statistical
mechanics and reflects the Gibbs-Boltzmann equilibrium ensemble.
The probability of energy increasing jumps is given by $\frac{1}{\lambda_2(t)}$,
with $\lambda_2(t)$ increasing in $t$.
This strategy  is designed to model an
initially hot system ($\lambda_2(0)$ small) with high probability of positive moves, which is
gradually quenched (the higher the temperature, the
more likely are moves upwards and viceversa); when the system is cool ($\lambda_2(t)$ large),
positive fluctuations are absent and the decreasing trajectories are forced to follow greedy-like
paths (very large jumps deep into a valley). With this type of algorithm
we try to take advantage during the paths both of the greedy-like behavior
and of the reluctant one (very small jumps and slow convergence). Obviously
the performance of the algorithm in terms of lowest energy found and execution time,
depends on the choice of the optimal cooling rate $\lambda_2(t)$. In \cite{CGGV2} the cooling rate,
with an exponential dependence on the time $\lambda_2(t)\propto k^{-t}$, was optimized for the SK model.

\begin{figure}
    \setlength{\unitlength}{1cm}
          \centering
               \includegraphics[width=8cm,height=8cm]{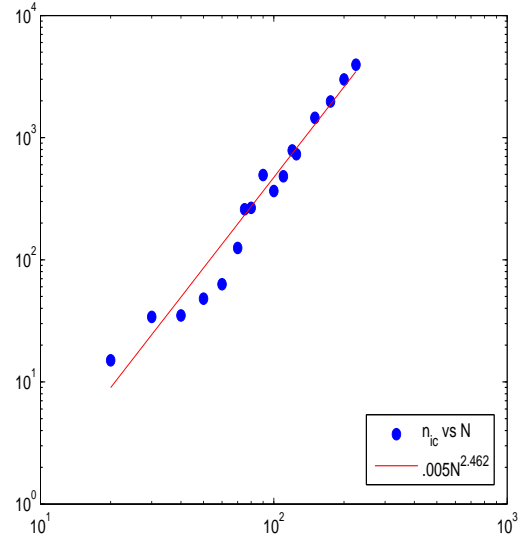}
               \caption{Scaling of the number of initial conditions per sample as a
               function of $N$ in log-log scale together with the best linear fit for
               the SK model.}\label{f:nci_N}
\end{figure}

The large statistics of ground state values required by our
analysis is obtained as follows. For a fixed disorder realization
$J$, we performed $n_{ic}$ independent runs of the algorithm. Each
run starts from an initial condition drawn at random from the
uniform distribution. The (approximate) ground state energy
$\en(J)$ is then obtained as the lowest energy of the metastable
states sampled along the $n_{ic}$ trajectories. With the above
strategy, the computed $\en(J)$ depends, in principle, on
$n_{ic}$. Therefore, the reliability of the results can be tested,
in a self consistent way, by computing the lowest energy from
larger and larger sets of initial conditions and by measuring the
number of runs $n_{ic}$ needed to stabilize the value within a
given accuracy $\delta$. This stabilized value is assumed to be
the ground state energy for the given realization of the model. In
the case of SK, for instance, running the algorithm with the
optimal cooling rate $\lambda_2(t)$ \cite{CGGV2}, the dependence
of $n_{ic}$ on the volume is $n_{ic}\simeq 0.005\ N^{2.462}$ for
$\delta=10^{-8}$ (see Fig.\ref{f:nci_N}). The linear system sizes,
the number of initial conditions $n_{ic}$ and the number of
disorder realizations $n_{J}$ used in our simulations are reported
in Tab. \ref{t:taPar} for both SK and EA model.
\begin{table}[thb]
\begin{center}
\begin{tabular}{|c|c|c|c|}\hline
    &$N$ &  $n_J$   &  $n_{ic}$\\ \hline
    & 50 & $2\cdot 10^5$   &     50 \\ \cline{2-4}
    & 75 & $1.5\cdot 10^5$ &    300 \\ \cline{2-4}
SK  &100 & $1.5\cdot 10^5$ &    500 \\ \cline{2-4}
    &150 & $2\cdot 10^4$   & $1.5\cdot 10^3$ \\ \cline{2-4}
    &200 & $1\cdot 10^4$   & $3\cdot 10^3$ \\ \hline \hline
    &$L$ & $n_J$   &  $n_{ic}$\\ \hline
    & 3-4 & $10^6$  &    200 \\ \cline{2-4}
    &  5  &  $10^5$  &    200 \\ \cline{2-4}
    &  6  &  $5\cdot 10^4$   &  $10^3$ \\ \cline{2-4}
EA$_{2D}$&  7  &  $5\cdot 10^4$   &  $10^4$  \\ \cline{2-4}
    & 8-9 &  $2\cdot 10^4$   &  $10^4$ \\ \cline{2-4}
    & 10  &  $2\cdot 10^4$  &   $2\cdot 10^4$ \\ \cline{2-4}
    & 11  &  $10^4$   &  $5\cdot 10^4$ \\ \hline\hline
    & 3-4 & $10^4$  &   $5\cdot 10^3$ \\ \cline{2-4}
EA$_{3D}$&  5  & $10^4$  &   $2\cdot 10^4$ \\ \cline{2-4}
    &  6  &  $2\cdot 10^3$   &  $10^5$ \\ \hline
\end{tabular}
\caption{Parameters of the simulations: linear system size,
number of disorder realizations $n_J$ and
number of initial conditions $n_{ic}$}\label{t:taPar}
\end{center}
\end{table}

\section{Results}
\vskip .6cm
We tested the power law (\ref{e:ScalEn}) for the mean ground state energy density $\epsn$
of the SK model,
sampling $17$ different system volumes $N$ between $20$ and $225$, with $n_J$ from $10^4$ up to
$2\cdot 10^5$ disorder realizations each (see Tab. \ref{t:taPar}).
These values of $n_J$
have been considered appropriate because of the (relatively) fast convergence of the sample mean
of $\en(J)$'s to its limit value, as $n_J$ is increased.

\begin{figure}
    \setlength{\unitlength}{1cm}
          \centering
               \includegraphics[width=8cm,height=8cm]{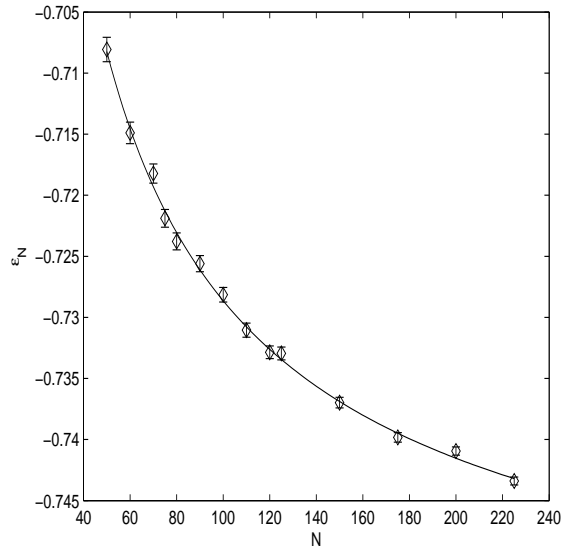}
               \caption{Mean ground state energy density $\epsn$ as a function of the volume $N$, together with
               the best three parameter numerical fit $\einf+bN^{-\omega}$ for the SK model.
               We obtain $\einf = -.7637(\pm 8\cdot 10^{-3})$, $b=.7383 (\pm 3.44\cdot 10^{-1})$ and
               $\omega=.6617 (\pm 1.52\cdot 10^{-1})$}\label{f:fit}
\end{figure}

Fitting the data $(N,\epsn)$ to the
power law  $\einf+bN^{-\omega}$ with three free parameters, we obtain the values
$\einf=-0.7627 (\pm 2.9 \cdot 10^{-3}), b = 0.7875 (\pm 7.95\cdot 10^{-2}), \omega=0.6820
(\pm 4.04\cdot 10^{-2})$.
The influence of the subleading corrections neglected in (\ref{e:ScalEn})
can be appreciated by restricting the fit to larger values of $N$.
Indeed, with $N\ge 50$ we have: $ \einf=-0.7637(\pm 8\cdot 10^{-3}),b=0.7383 (\pm 3.44\cdot 10^{-1}),
\omega = 0.6617 (\pm 1.52\cdot 10^{-1}) $, (see Fig.~\ref{f:fit}). In both cases, the estimated value of $\einf$
is in good agreement with
Parisi's analytical value $\einf^P=-0.7633$ \cite{MPV}. The scaling exponent
 has been computed in previous numerical studies by Palassini \cite{P}, Bouchaud et al. \cite{BKM},
Katzgraber and Campbell \cite{KC};
all these results suggest $\omega=\frac 2 3$.
 Our estimate of $\omega$ with $N\ge 50$ is compatible with this value, even though
the uncertainty on the data are much larger than those presented in \cite{BKM}. On the other hand, the error on
the exponent can be lowered by fitting the data with only two free parameters and letting $\einf=\einf^P$;
in doing so, we obtain $\omega=-0.6685\pm 2.23\cdot 10^{-2}$ (for $N\ge 50$) and $\omega=-0.6738\pm 9\cdot 10^{-3}$
for $N\ge 20$. Moreover, fixing also $\omega=\frac 2 3$ we obtain $b=0.7514\pm 4\cdot 10^{-3}$ for $N\ge 50$ and
$b=0.7475\pm 6\cdot 10^{-3}$ for $N\ge 20$.
Let us conclude this point by observing that the previous estimates of $b$ are quite close to the value of
 $A=0.77\pm 0.01$  obtained in \cite{ABMM} fitting $e_\infty+AN^{-\frac 23}$ to the values of the internal
energy at $T=0.4$. In \cite{ABMM} the power law $N^{-\frac 23}$ has been found to describe closely the data in the whole
spin glass phase.

\begin{figure}
    \setlength{\unitlength}{1cm}
          \centering
               \includegraphics[width=8cm,height=8cm]{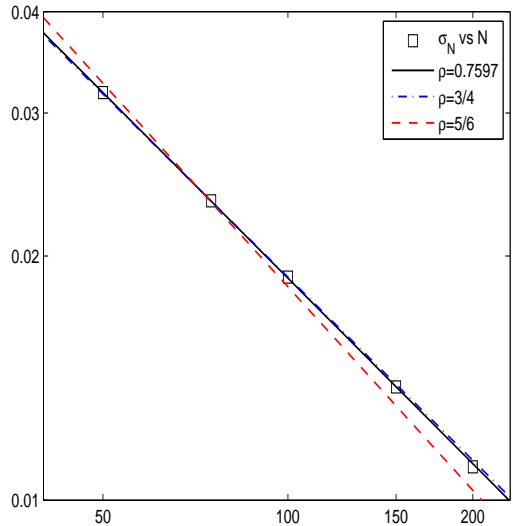}

               \caption{Standard deviation $\sigma_N$ of the  ground state
               energy as a function of $N$ in log-log scale for the SK model. The lines represent
               the power law $aN^{-\rho}$ for the three values of $\rho$: $3/4$, $5/6$ and $0.7597$
               (the last one is obtained with a two parameter fit)}\label{f:fitsig}
\end{figure}

\begin{figure}
    \setlength{\unitlength}{1cm}
          \centering
               \includegraphics[width=8cm,height=8cm]{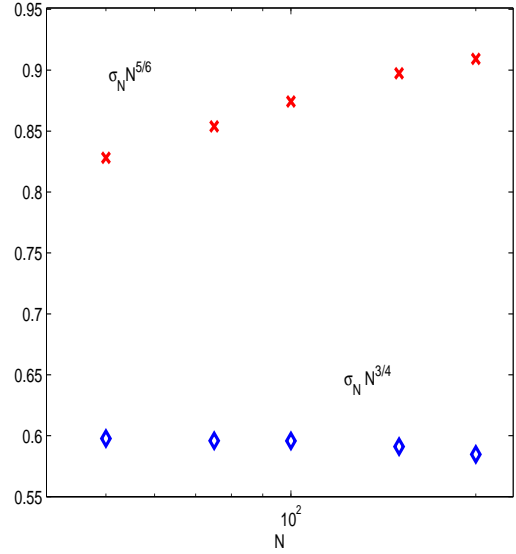}
               \caption{Plots of $\sigma_NN^{3/4}$ and $\sigma_NN^{5/6}$.}\label{f:S_devtot}
\end{figure}

In Fig.~\ref{f:fitsig} we represent in log-log scale the numerical
data for $\sigma_N$ as a function of $N$ together with the
best numerical fit $\sigma_N=aN^{-\rho}$, which gives
$\rho=-.7597\pm 1.04\cdot 10^{-2}$.
As already said in the introduction, in literature two values are conjectured for the scaling exponent
of the sample to sample fluctuations: $\rho=3/4$ and $\rho=5/6$ \cite{BKM,P,AMY,CMPP}.
As it is apparent from Figg.~\ref{f:fitsig} and \ref{f:S_devtot},
the exponent $\rho=3/4$ (blue line) is consistent with our data, while
$\rho=5/6$ seems to be less likely to occur (red line in Fig.~\ref{f:fitsig} and
red crosses in Fig.~\ref{f:S_devtot}). This is in agreement with the results in
literature in which all the estimates are smaller than $5/6$. In particular, the value $\rho=3/4$ was obtained
in \cite{P} and later confirmed in \cite{BKM}, but while in the former paper $\rho=5/6$ is ruled out,
the authors of the latter can not draw the same conclusion. A similar uncertainty is present also in simulations
at finite temperature; indeed, in \cite{ABMM} at $T=0.4$ the data for the internal energy are compatible with
$\rho=5/6$ but  $\rho=3/4$ is not ruled out.
The value $\rho=3/4$ is supported also by some recent theoretical results \cite{PR,ABMM}
suggesting that the sample-to-sample fluctuations should be proportional to $N^{-5/6}$.
\par
\begin{figure}
    \setlength{\unitlength}{1cm}
    \begin{minipage}[t]{7cm}
          \centering
               \includegraphics[width=7cm,height=7cm]{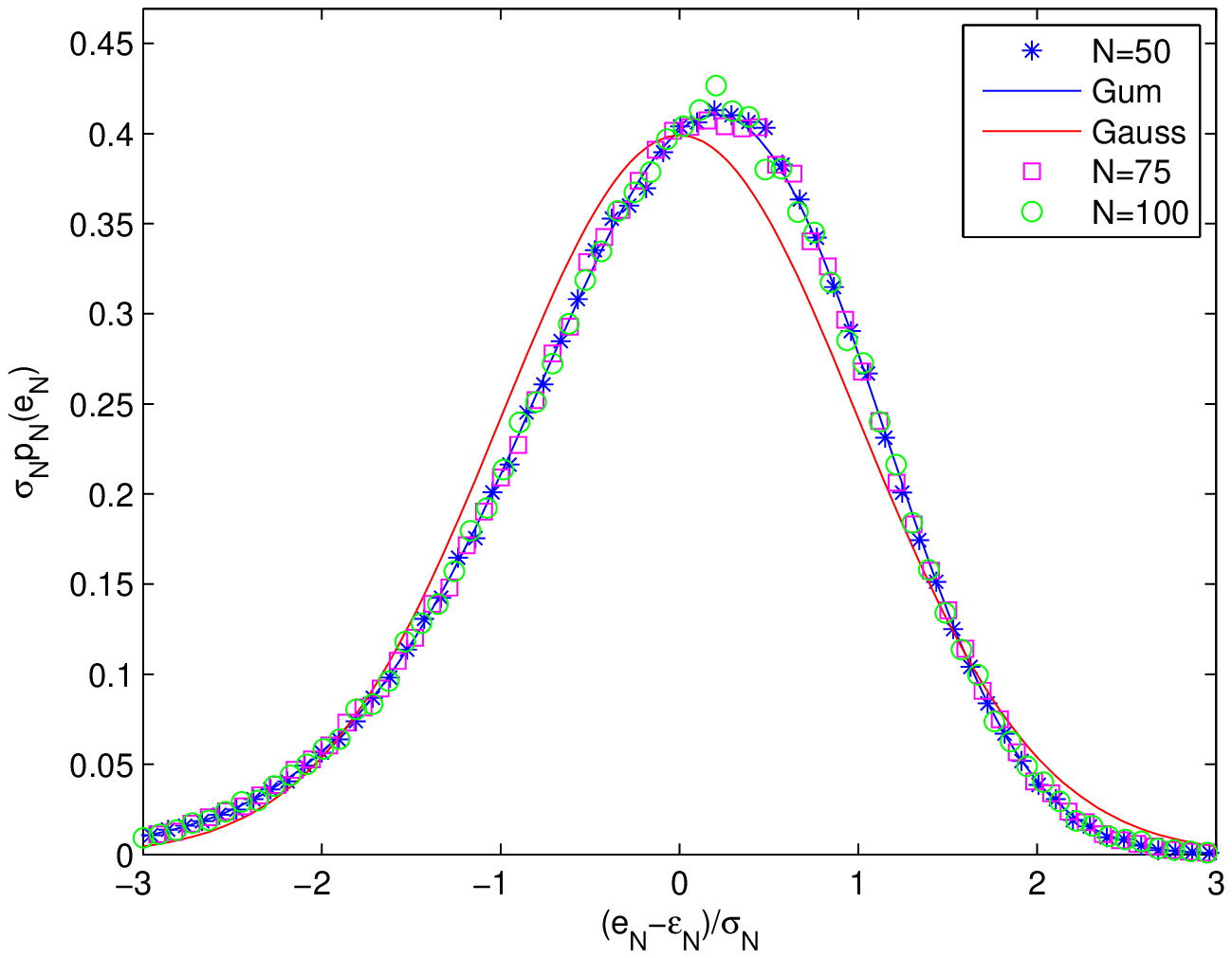}
               \caption{Scaling plot of the ground state energy PDF.
               The dotted line is the Gumbel PDF with $m=6$ ($u=.2011219,
               v=2.348408, w=165.5589$). All PDF ($N=50,75,100$) are normalized to one and
               have zero mean and unit variance.}\label{f:Prob}
   \end{minipage}
   \ \hspace{.3cm} 
   \begin{minipage}[t]{7cm}
         \centering
               \includegraphics[width=7cm,height=7cm]{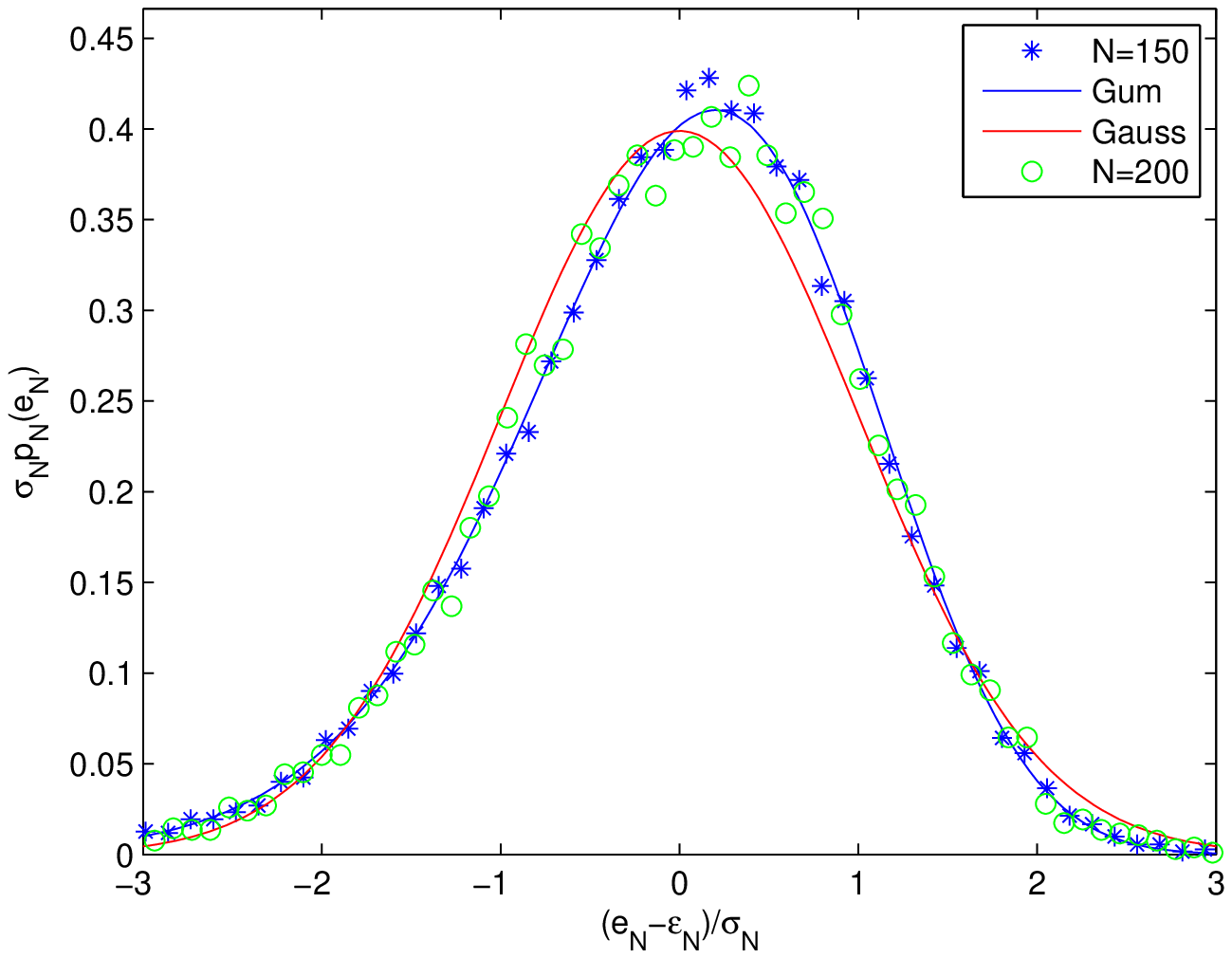}
               \caption{Scaling plot of the ground state energy PDF.
               The dotted line is the Gumbel PDF with $m=6$ ($u=.2011219,
               v=2.348408, w=165.5589$). All PDF ($N=150,200$) are normalized to one and
               have zero mean and unit variance.}\label{f:Prob1}
   \end{minipage}
\end{figure}

The detection of the distribution $p_N(e_N)$ for large $N$ is obviously much more difficult
than studying the two moments $\epsn$ and $\sigma_N^2$. Such a study has been addressed in \cite{BKM,P,ABMM}
via numerical simulations; here we will give some insight into this issue following closely the approach of \cite{P}.
The problem pertains to the  statistics of extremes; in this theory the scaling of the minimum of a family
of $M$ random  variables is studied for large $M$. Here the random variables are the energies $H_N(J,\ve{S})$ of
the $M=2^N$ spin configurations $\ve{S}$, the minimum is the ground state energy $\en(J)$, and the scaling is studied
using the variables $x_N=(\en-\epsn)/\sigma_N$. In the quoted papers the data are tested against the standard
extreme values distributions of a family of identically distributed (i.i.d.) random variables: 1) Gumbel,
if the individual
distribution of the random variables is unbounded and decreases faster than any power law;
2) Fisher, if the  distribution decreases as a
power law; 3)Weibull, if the  distribution has a cut-off. In the papers \cite{BKM,P} it is shown that none
of the previous distributions describe the data. The same negative result was obtained \cite{P,ABMM}
testing the Tracey-Widom distribution for correlated variable.
While it is not unexpected the failure of the Gumbel distribution in describing the data
(since the energy levels that contribute to the ground state energy for the mean field SK model are
not independent), it is much more surprising that the approximate behavior of the data can be found in the
family of the generalized Gumbel distributions.
These are the distributions the $m$-th smallest value in a set of i.i.d. random variables \cite{Gu}
$$
g_m(x)=w \exp \left[m\frac{x-u}{v}-m\exp \frac{x-u}{v}\right],
$$
where $u$, $v$ and $w$ are constant parameter.
In his paper \cite{P} Palassini found that the Gumbel distribution with $m=6$ describes quite closely the
ground state energy distribution of SK; our data support this statement. In Fig.~\ref{f:Prob} and \ref{f:Prob1}
the empirical distributions of $x_N$ for the volumes $N=50,75,100,150$, $200$ are reported together with $g_6(x)$
and the standard normal distribution. The values of parameters in $g_6(x)$, taken from \cite{P}, are chosen imposing
zero mean and unit variance. Even if the data for the larger sizes, see Fig.\ref{f:Prob1}, are much more
noisy than those for the smaller ones, the agreement with $g_6(x)$ seems fairly clear for all the sizes.
A further support can be obtained by comparing the curtosis and the skewness of $g_6(x)$ with the same cumulants
of the variable $x_N$ at different sizes, see table \ref{t:tabKur}.

\begin{table}[thb]

\begin{center}
\begin{tabular}{|c|c|c|}\hline
$N$ & Kurtosis &  Skewness \\ \hline
 50 & 3.4249   &  -.4451   \\ \hline
 75 & 3.3148   &  -.4052   \\ \hline
100 & 3.3464   &  -.4046   \\ \hline
150 & 3.5515   &  -.4346   \\ \hline
200 & 3.3312   &  -.3599   \\ \hline
$g_6(x)$ & 3.3535 &-.4247  \\ \hline
\end{tabular}
\caption{Kurtosis and skewness for SK model}\label{t:tabKur}
\end{center}
\end{table}

The second family of spin system we study is the Edwards-Anderson
model. As we said in the introduction, this should be considered as
a preliminary study of the behavior of the algorithm in exploring
the low energy configurations of the finite dimensional spin glass
model. Indeed, we use the cooling rate optimized for the SK model,
both for the two-dimensional and three dimensional cases (square and
cubic lattices of linear size $L$). However, while in the three
dimensional case the problem of finding the  ground states is
NP-hard, the two dimensional one is polynomial \cite{Barahona}. For
this reason the three dimensional case is only partially studied,
i.e. only small linear sizes are considered.

Nevertheless, our data for $3d$-model confirm the results present in the literature \cite{BKM}.
All the parameters used in the numerical simulations both for two dimensional and three
dimensional EA model are listed in Tab. \ref{t:taPar}.

\begin{figure}
    \setlength{\unitlength}{1cm}
          \centering
               \includegraphics[width=8cm,height=8cm]{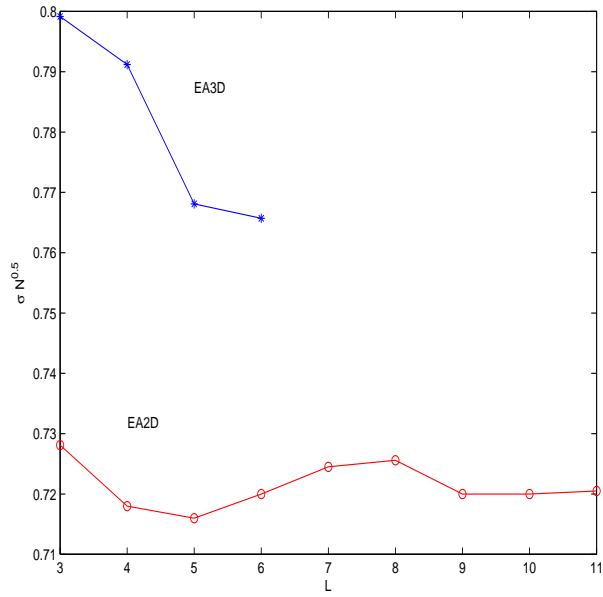}
               \caption{Standard deviation of the ground state energy times
               the square root of the volume for the EA model as a function of $L$.}\label{f:sigEA}
\end{figure}

\begin{figure}
    \setlength{\unitlength}{1cm}
          \centering
               \includegraphics[width=8cm,height=8cm]{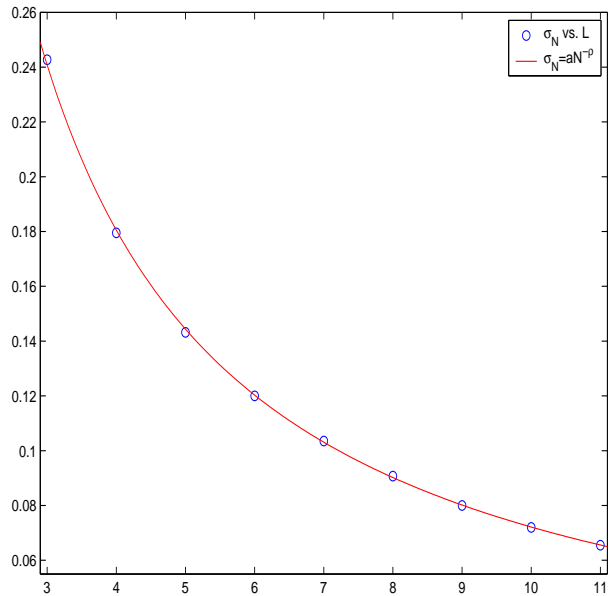}
               \caption{Best fit of the standard deviation of the ground state energy
               (intensive quantity) for the EA model ($d=2$) as a function of $L$,
               which gives $\rho=0.501\pm 6\cdot 10^{-3}$.}\label{f:fitsigEA}
\end{figure}

In order to validate our numerical experiment, it is useful to
compare the numerical data for $\sigma_N$ with the theoretical prediction $\rho=\frac 1 2$.
In Fig.~\ref{f:sigEA} we show
$\sigma_NN^{\rho}\equiv\sigma_N\sqrt{N}$; while for $d=2$ our data
are compatible with size independence, for $d=3$ a first sketchy study
seems to show no analogous trend. In Fig.~\ref{f:fitsigEA} we represent (for $d=2$)
the best numerical fit $aN^{-\rho}$ which gives $\rho=0.501\pm 6\cdot 10^{-3}$
(and $a=.731\pm 1.4\cdot 10^{-2}$) in accordance with the expected value.
This confirms the validity of our algorithm
and the consistency of our numerical results for this model.
Such a scaling low is expected for the central limit theorem in the cases in
which the different terms contributing to the ground-state energy are
independent. As a matter of fact, for EA model these terms are
correlated; so this type of scaling reads as an indication of weak correlation.
In contrast, for SK model, the variables that play a role in the ground state energy
are sufficiently correlated to prevent the central limit theorem behavior.


\begin{figure}
    \setlength{\unitlength}{1cm}
          \centering
               \includegraphics[width=12cm,height=12cm]{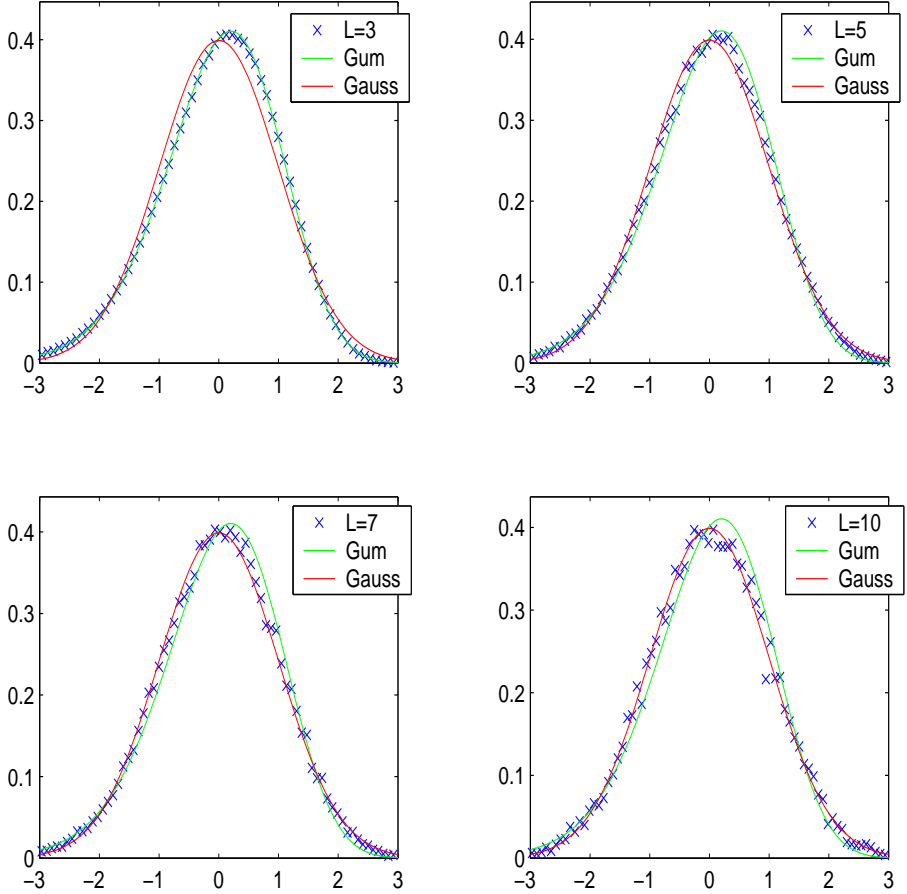}
               \caption{Scaling plot of the ground state energy PDF for the EA model ($d=2$).
               The green line is the Gumbel PDF with $m=6$. The red line is
               a Gaussian PDF.}\label{f:ProbEA}
\end{figure}

A further support to the hypothesis of weak correlation comes from the study
of the limiting shape of the distribution  $p_N(\en)$.
Because of the dependence of the random variables $\en(J)$,
we can not have a Gaussian shape for $p_N(\en)$ at large $L$ (see for example \cite{WA}),
even though a weakly Gaussian-like  behavior is expected
(it follows from the Brout's heuristic argument \cite{BiYo} and also from replica theory calculations \cite{AspMo}).
In fact, in the case of the two dimensional lattice,  the numerical data for small volumes show
that the PDF follows the same Gumbel distribution (with $m=6$) already found
as the limiting behavior of the SK model. However, as $L$
is increased, we observe that the distribution moves away from the Gumbel  and
seems to approach a Gaussian limiting shape
(see Fig.~\ref{f:ProbEA}).

\begin{figure}
    \setlength{\unitlength}{1cm}
    \begin{minipage}[t]{7cm}
          \centering
               \includegraphics[width=7cm,height=7cm]{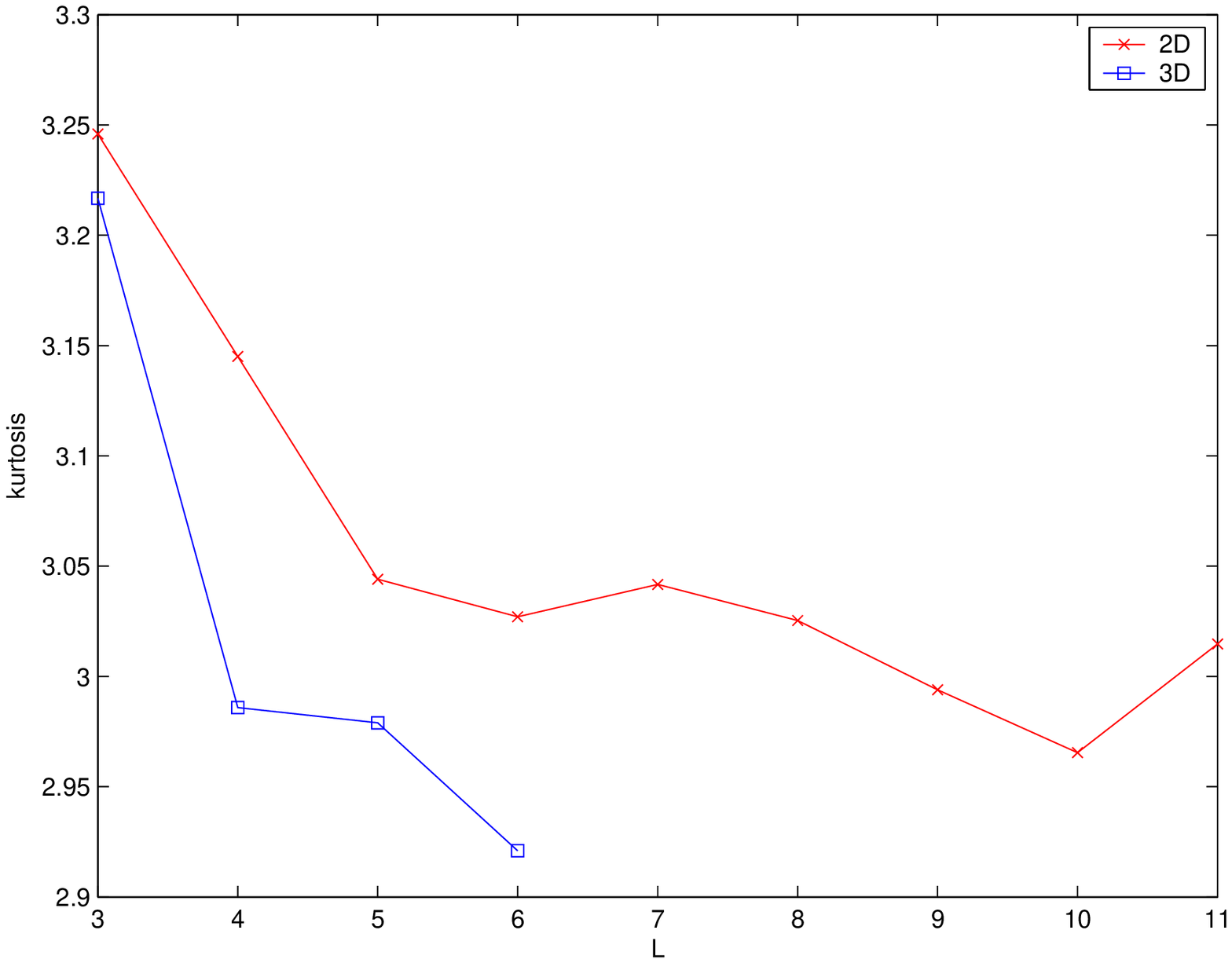}
               \caption{Kurtosis of the ground state energy distribution
               for EA model ($d=2$ and $d=3$) as a function of $L$.}\label{f:Kur}
   \end{minipage}
   \ \hspace{.3cm} 
   \begin{minipage}[t]{7cm}
         \centering
               \includegraphics[width=7cm,height=7cm]{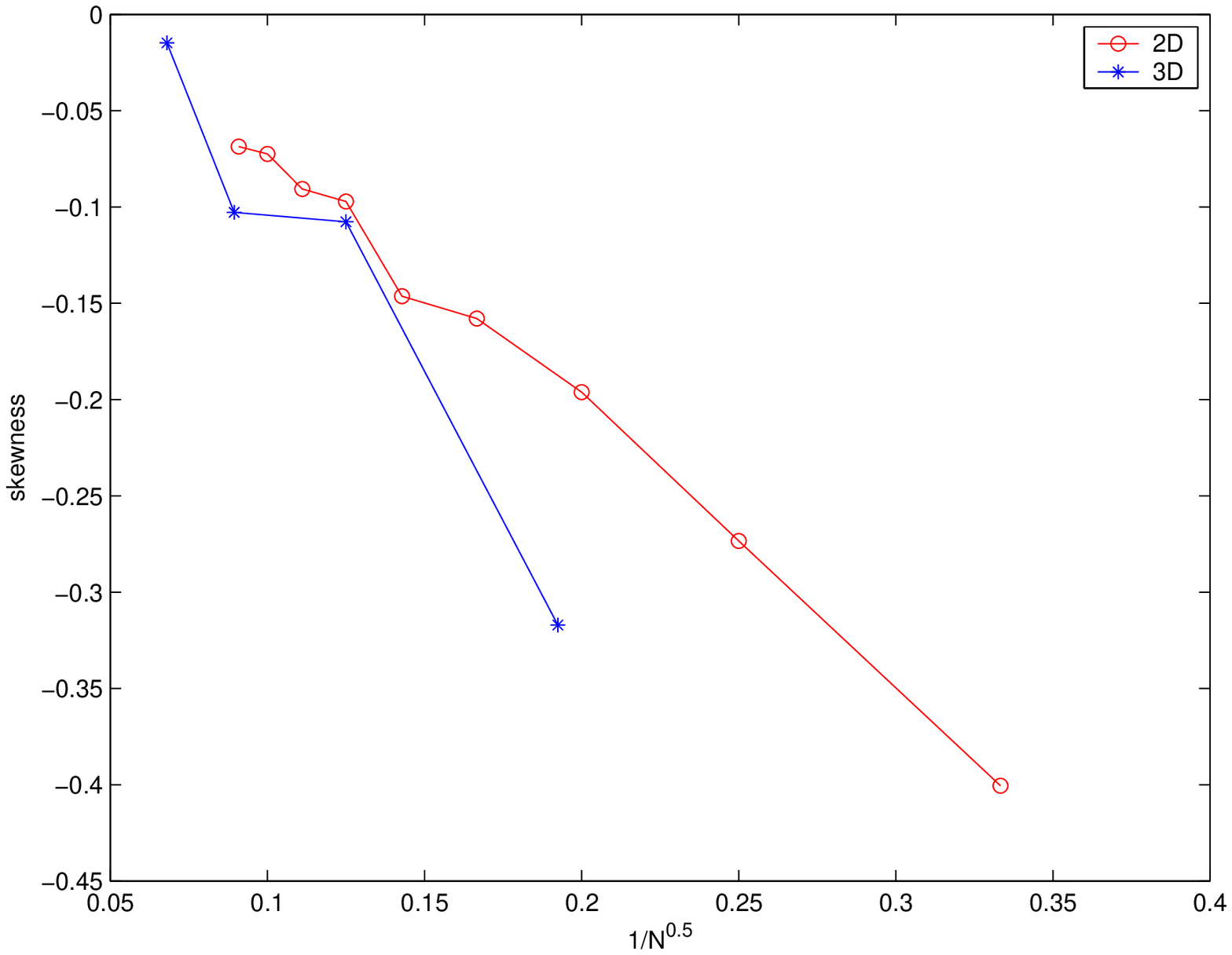}
               \caption{Skewness of the ground state energy distribution
               for EA model ($d=2$ and $d=3$) as a function of $1/N^{0.5}$.}\label{f:Skew}
   \end{minipage}
\end{figure}

If we consider the kurtosis and the skewness of $p_N(\en)$ we have that they decay with
the system size (Fig.~\ref{f:Kur} and \ref{f:Skew}). The data in Fig.~\ref{f:Skew}
suggest zero limiting values as $N\to\infty$ in accordance
with the central limit theorem law that predicts that the skewness scales as $N^{-1/2}$
(in Fig.~\ref{f:Skew} the data are completely compatible with a linear convergence to zero, confirming
a central limit-like scaling).

\begin{figure}
    \setlength{\unitlength}{1cm}
          \centering
               \includegraphics[width=8cm,height=8cm]{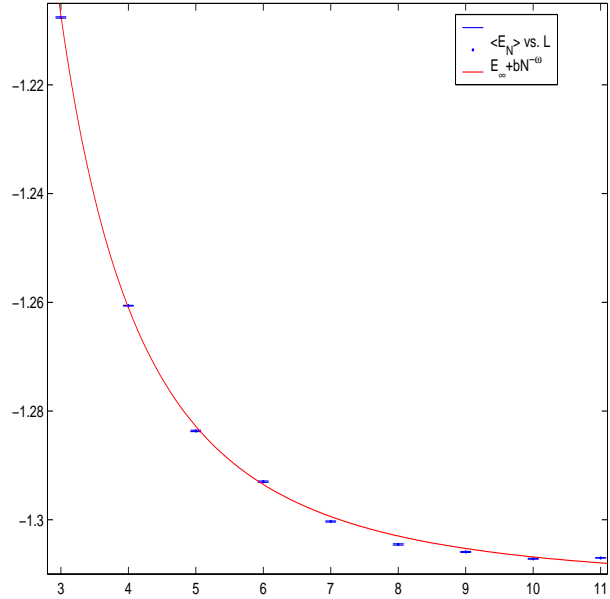}
               \caption{Mean ground state energy density with errorbars and the best fit
               (intensive quantity) for the EA model ($d=2$) as a function of $L$.}\label{f:fiteminEA}
\end{figure}

\begin{figure}
    \setlength{\unitlength}{1cm}
       \begin{minipage}[t]{7cm}
         \centering
               \includegraphics[width=7cm,height=7cm]{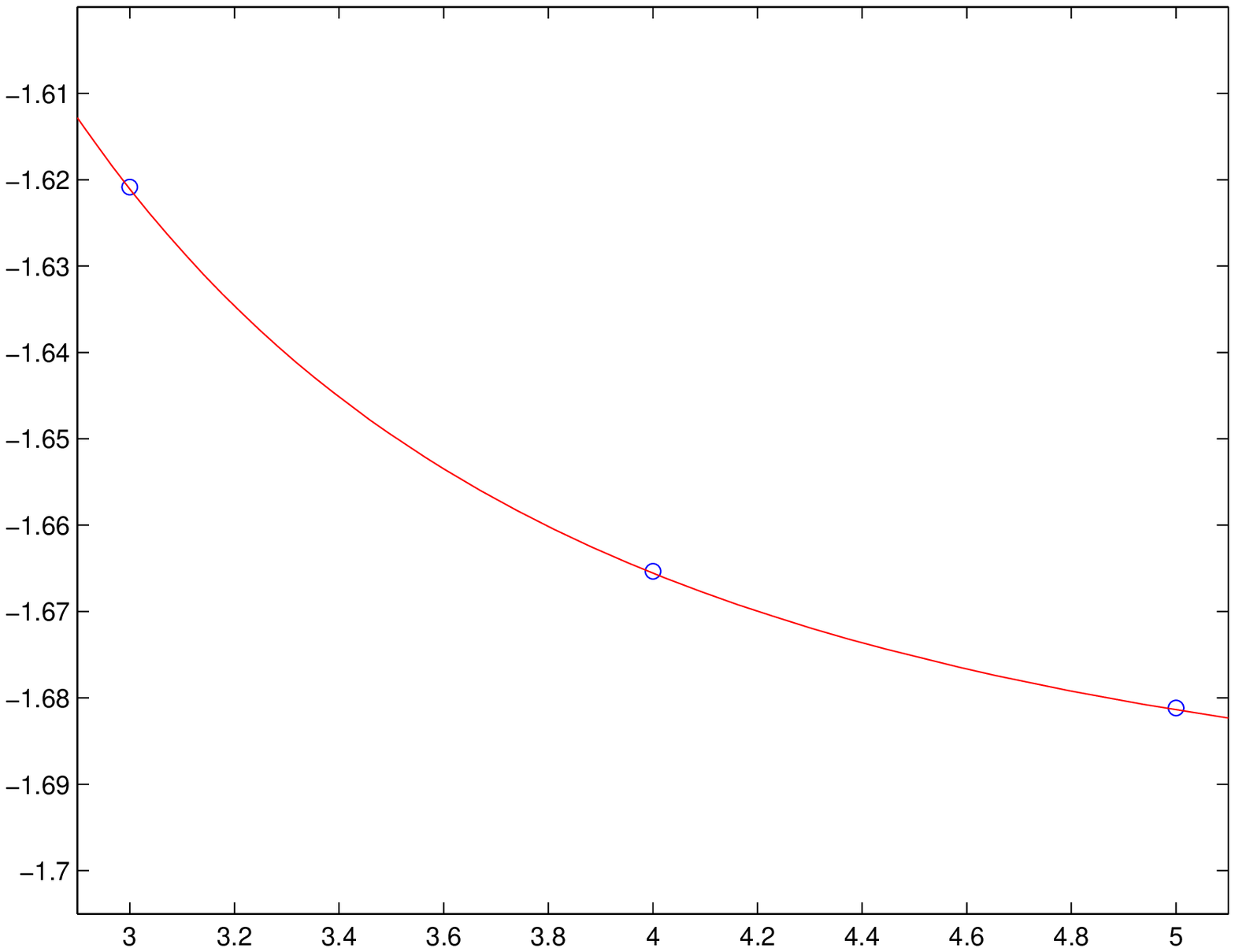}
               \caption{Mean ground state energy density and the best fit
               (intensive quantity) for the EA model ($d=3$) as a function of $L$.}\label{f:fit3d}
   \end{minipage}
      \ \hspace{.3cm} 
    \begin{minipage}[t]{7cm}
          \centering
               \includegraphics[width=7cm,height=7cm]{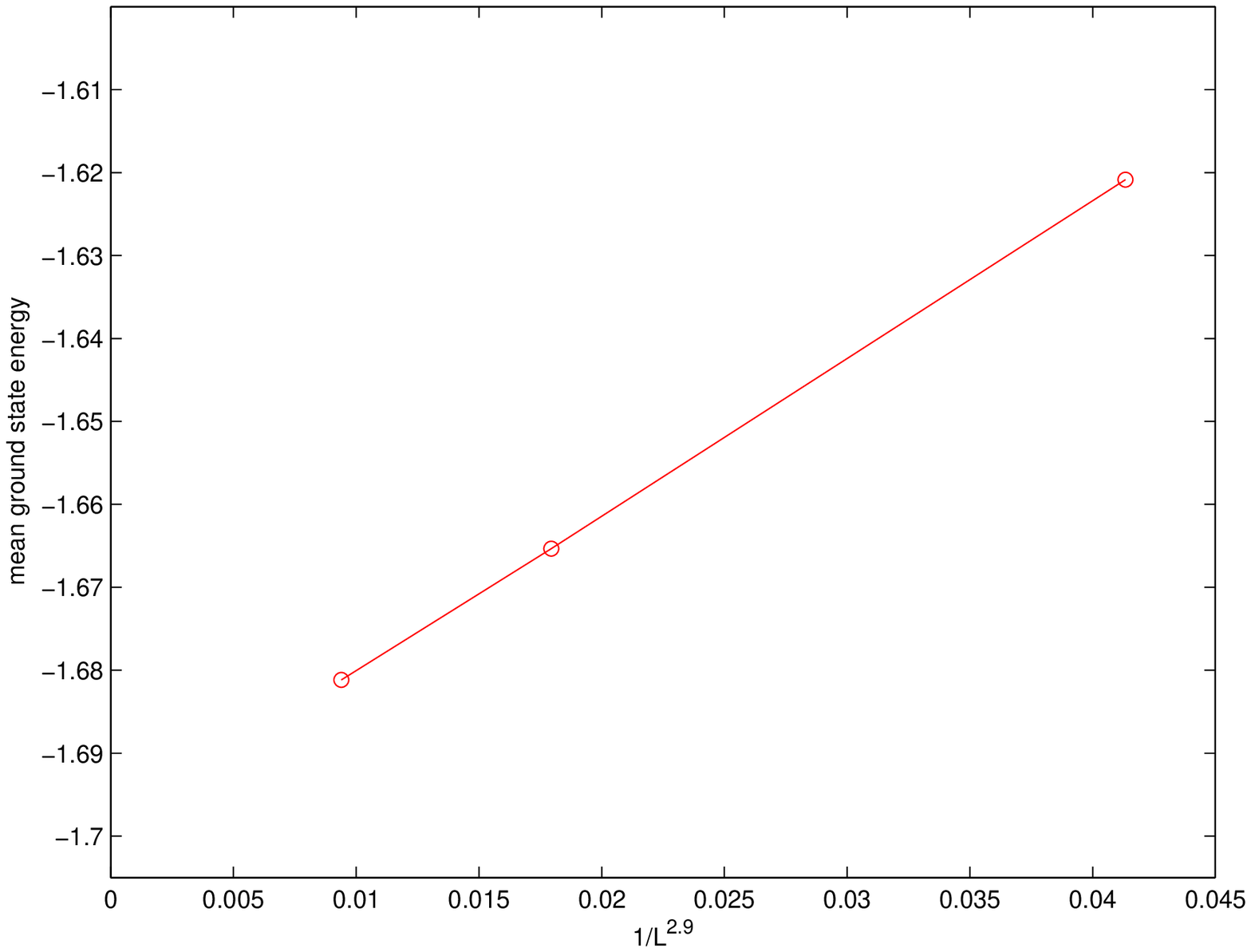}
               \caption{Mean ground state energy
               for EA model ($d=3$) as a function of $L^{-2.9}$.}\label{f:fitl3d}
   \end{minipage}

\end{figure}

Now we move to the study of the finite-size corrections to the mean energy density
which, to the leading order, are expected to scale as $N^{-\omega}$.
For EA model the analytic knowledge of $\epsn$ in the
thermodynamic limit is not available (unlike in the case of SK model).
First we consider the case $d=2$. We fit our data ($3\le L\le 11$)
to Eq.~\ref{e:ScalEn} where $\einf$, $b$ and
$\omega$ are free parameters. We obtain (Fig.~\ref{f:fiteminEA}):
$\einf = -1.312 (\pm 2\cdot 10^{-3})$, $b = 1.637 (\pm 2.94\cdot 10^{-1})$
and $\omega = 1.251 (\pm 8.5\cdot 10^{-2})$
in accordance with \cite{BKM} that provides $\omega=1.175\pm 5\cdot 10^{-3}$.
For $d=3$ the best fit gives a mean (intensive) ground state energy decaying as
$-1.698+2.077N^{-1}$ (see Fig~\ref{f:fit3d}).
In \cite{BKM} the estimated value of $\omega$ is $0.967\pm 1\cdot 10^{-2}$, even though, due to
the uncertainty on the data, any value between $0.933$ and $1$ seems possible.
In order to check the consistency of our data with this value we compute the linear fit $\einf+bN^{-0.967}$,
represented  in Fig.~\ref{f:fitl3d},
which gives $\einf = -1.699 (\pm 3\cdot 10^{-3})$, $b = 1.891 (\pm 1.32\cdot 10^{-1})$,
in good agreement with the values presented in \cite{BKM}.

To conclude, we presented a numerical investigations on mean field and finite-dimensional
spin glass models that produce finite size scaling exponents consistent
with the results of many works on the subject, but produced with different algorithms.
The study of the probability distribution of the ground state energy
showed that in the large volume limit it becomes Gaussian for the EA model, but with a
finite size behavior close to a Gumbel distribution which, on the other hand, is the limiting
distribution for the SK model. Further effort is necessary to produce the optimal algorithm in
searching for the ground state of the three-dimensional EA model.

\vskip .5cm
{\bf Acknowledgments}. We thank Cristian Giardin\`a and Pierluigi
Contucci for many useful discussions that stimulated and improved
this work.

\end{document}